\documentclass[Physsubmission, Phys]{SciPost}

\hypersetup{
    colorlinks,
    linkcolor={red!50!black},
    citecolor={blue!50!black},
    urlcolor={blue!80!black}
}

\usepackage[bitstream-charter]{mathdesign}
\DeclareSymbolFont{usualmathcal}{OMS}{cmsy}{m}{n}
\DeclareSymbolFontAlphabet{\mathcal}{usualmathcal}

\newcommand{\braque}[1]{{\langle #1 \rangle}}
\newcommand{\ket}[1]{{\vert #1 \rangle}}

\newcommand{\lbl}[1]{\label{eq:#1}}

\newcommand{\lblfig}[1]{\label{fig:#1}}

\newcommand{\rf}[1]{(\ref{eq:#1})}
\newcommand{\fig}[1]{(\ref{fig:#1})}
\newcommand{\be}{\begin{equation}}
\newcommand{\en}{\end{equation}}
\newcommand{\ba}{\begin{array}}
\newcommand{\ea}{\end{array}}
\newcommand{\bp}{\begin{pmatrix}}
\newcommand{\ep}{\end{pmatrix}}
\newcommand{\bc}{\begin{center}}
\newcommand{\ec}{\end{center}}
\newcommand{\bt}{\begin{tabular}}
\newcommand{\et}{\end{tabular}}
\newcommand{\mpi }{m_\pi}
\newcommand{\meta }{m_\eta}
\newcommand{\mpid}{{m^2_{\pi}}}
\newcommand{\metad }{m^2_{\eta}}
\newcommand{\mK }{m_K}
\newcommand{\mkd}{{m^2_{K^+}}}
\newcommand{\pip}{\pi^+}
\newcommand{\pim}{\pi^-}
\newcommand{\piz}{\pi^0}
\newcommand{\Kp}{{K^+}}
\newcommand{\Km}{{K^-}}
\newcommand{\Kz}{{K^0}}
\newcommand{\Kzb}{\bar{K}^0}
\newcommand{\Kbar}{\bar{K}}
\newcommand{\qd}{{q^2}}
\newcommand{\im}{{\rm Im\,}}
\newcommand{\disc}{{\rm disc\,}}
\newcommand{\lapprox}{%
\mathrel{%
\setbox0=\hbox{$<$}\raise0.6ex\copy0\kern-\wd0\lower0.65ex\hbox{$\sim$}}}
\renewcommand{\slash}[1]{%
\mathrel{\setbox0=\hbox{$/$}\copy0\kern-\wd0\hbox{$#1$}}}

\begin{document}

\begin{center}{\Large \textbf{
Deriving experimental constraints on the scalar form factor in the second-class
$\tau \to\eta \pi \nu$ mode
}}\end{center}

\begin{center}
B. Moussallam\textsuperscript{1}  
\end{center}

\begin{center}
{\bf 1} Laboratoire Ir\`ene Joliot-Curie, Universit\'e
Paris-Saclay, 91406 Orsay, France
\end{center}

\begin{center}
\today
\end{center}


\definecolor{palegray}{gray}{0.95}
\begin{center}
\colorbox{palegray}{
  \begin{minipage}{0.95\textwidth}
    \begin{center}
    {\it  16th International Workshop on Tau Lepton Physics (TAU2021),}\\
    {\it September 27 – October 1, 2021} \\
    \doi{10.21468/SciPostPhysProc.?}\\
    \end{center}
  \end{minipage}
}
\end{center}

\section*{Abstract}
{\bf
The rare second-class decay mode of the $\tau$ into $\eta\pi\nu$ could be
observed for the first time at Belle II. It is important to try to derive a
reliable evaluation of the branching fraction and of the energy distribution
of this mode within the standard-model. Many predictions exist already in the
literature which can differ by one to two orders of
magnitude. Here, an approach based on a systematic use of the
property of analyticity of form factors and scattering amplitudes in
QCD is discussed. In particular, we will show that the scalar form factor
in the $\tau$ decay can be related to photon-photon scattering and
radiative $\phi$ decay amplitudes for which precise experimental
measurements have been performed by the Belle and KLOE collaborations.
}



\section{Introduction}
The decay $\tau\to \eta\pi\nu$ is induced by second-class type
currents~\cite{Weinberg:1958ut}. In the Standard Model (SM) the amplitude must
thus be proportional to the small isospin breaking parameters:
$(m_d-m_u)/m_s$, $e^2$ and it has a sensitivity to new scalar or tensor
interactions (e.g.~\cite{Bramon:1987zb,Nussinov:2008gx,Garces:2017jpz}).  
As Tony Pich has reminded us in the introductory talk, this mode was
once claimed to have been observed~\cite{Derrick:1987sp} with an
unexpectedly large branching fraction $BF=(5.1\pm1.5)\%$ but this was
not confirmed.  The best upper bounds now available are $BF <
7.3\times10^{-5}$ (Belle~\cite{Hayasaka:2009zz}) and $BF <
9.9\times10^{-5}$ (Babar~\cite{delAmoSanchez:2010pc}). At the new
Belle II facility, it has been estimated~\cite{Ogawa:2020iwi} that the
$\tau\to \eta\pi\nu$ decay could be observed with a significance of
$2.6\sigma$ for a branching fraction of $4.4\times10^{-5}$.
A second-class scalar current would lead to a simple modification
of the scalar form factor (defined below~\rf{ffactorsdef}) by a term
linear in energy
\be
f_0^{\eta\pi}(s) = \left. f_0^{\eta\pi}(s)\right\vert_{SM}\left(1 -\epsilon_S
\frac{s}{m_\tau(m_d-m_u)}\right)
\en
proportional to the strength $\epsilon_S$ (with the notation
of~\cite{Bhattacharya:2011qm}) of the new interaction. A correct
estimate of the scalar form factor in the SM is necessary in order to
derive reliable constraints on $\epsilon_S$ from experimental results.

A number of evaluations of the $\eta\pi$ vector and scalar form factors have
already been performed
(e.g.~\cite{Tisserant:1982fc,Pich:1987qq,Neufeld:1994eg,Nussinov:2008gx,Paver:2010mz,Volkov:2012be,Descotes-Genon:2014tla,Escribano:2016ntp}
for a representative list).
The predictions for the branching fraction associated with the vector
form factor lie in a range $BF_V\simeq[0.1-0.8]\cdot10^{-5}$ while those
associated with the scalar form factor seem more uncertain: 
$BF_S\simeq[0.3-5.3]\cdot10^{-5}$. 

We present here an evaluation of $f_0^{\eta\pi}$ which exploits the
properties of analyticity and unitarity of form factors in QCD
(e.g.~\cite{Barton:1965}). In this framework, generating the energy
dependence of the form factors can be viewed as a final-state
interaction (FSI) problem. This allows us to derive relations between
the scalar form factor and the photon-photon production amplitude
$\gamma\gamma\to \eta\pi$ and also with the $\phi$ radiative decay amplitude
$\phi\to\gamma \eta\pi$, for which FSI theory can also be applied and
which are experimentally known.

\section{Unitarity and Omn\`es representations of form factors}
The $\tau$ decay amplitude into two pseudoscalar mesons involves the
matrix element of the charged vector current
$\braque{P_1(p_1)P_2(p_2)\vert j_\mu^{uq}(0)\vert 0}$ (with
$j_\mu^{uq}=\bar{u}\gamma_\mu{q}$, $q=d,s$), which can be expressed in
terms of two independent form factors
\be\lbl{ffactorsdef}
\braque{P_1(p_1)P_2(p_2)\vert \bar{u}\gamma^\mu{q}\vert 0}=C_{12}\Big\{
f_+^{P_1P_2}(s) \Big[p_1-p_2-\frac{\Delta_{12}}{s}(p_1+p_2)\Big]^\mu +
f_0^{P_1P_2}(s)\, \frac{\Delta_{12}}{s}[p_1+p_2]^\mu
\Big\}
\en
where $s=(p_1+p_2)^2$, $\Delta_{12}=m_1^2-m_2^2$ and $C_{12}$ is a
numerical factor (below, we will need $C_{\eta\pip}=-\sqrt2$,
$C_{\Kzb\Kp}=-1$). Using the Ward identity for the vector current, the
scalar form factors $f_0^{P_1P_2}$ can also be expressed in terms of the
matrix elements,
\be
C_{12}\Delta_{12} f_0^{P_1P_2}(s)=
-\braque{P_1(p_1)P_2(p_2)\vert (m_d-m_u)\bar{u}d-eA_\mu{j}^{\mu,ud}  \vert 0}\ .
\en

For the $\pi\pi$ or $\pi{K}$ vector form factors, simple approximate
evaluations are known to hold e.g.
\be\lbl{Kpiomnes}
f_+^{\pi K}(s)\simeq f_+^{\pi K}(0) \Omega_1(s),\quad
\Omega_1(s)=\exp\Big(\frac{s}{\pi}\int_{(\mpi+\mK)^2}^\infty ds'\,
\frac{\delta^{1/2}_1(s')}{s'(s'-s)}
\Big)
\en
where $\Omega_1$ is the Omn\`es function~\cite{Omnes:1958hv} and
$\delta^{1/2}_1$ is the $J=0$ $I=1/2$ $\pi{K}$ scattering phase
shift. This is justified because $\pi{K}$ scattering is effectively
elastic in a energy region $s < \Lambda^2$, with $\Lambda\simeq1$ GeV,
which includes the leading low energy resonance $K^*(892)$.
Multiplying $f_+^{\pi K}$ by the inverse of the Omn\`es function
removes the cut below $\Lambda$, such that in the region $s <
\Lambda^2$ one can write a low energy expansion: $\Omega^{-1}(s)
f_+^{\pi K}(s)= f_+^{\pi K}(0)(1+O(s/\Lambda^2))$.
The situation is different for the $\pi\eta$ vector
form factor because contributions to  unitarity  arise form the states
$\ket{\gamma\pip}$, $\ket{\piz\pip}$ which are lighter than the elastic
threshold  $\ket{\eta\pip}$. The contribution from the
$\ket{\piz\pip}$ should be the dominant one below 1 GeV as it
is enhanced by the $\rho(770)$ resonance. The corresponding
discontinuity reads, 
\be
\disc[f_+^{\pi\eta}(s)]_{\pi\pi}=2i
\theta(s-4\mpid)\frac{4\mpid-s}
   {\sqrt{\lambda_{\eta\pi}(s)}} F_V^\pi(s)\times( t_{J=1}^{\piz\pip\to \eta\pip}(s))^*
\en
where $\lambda_{12}(s)=(s-(m_1+m_2)^2)(s-(m_1-m_2)^2)$. It involves
the pion form factor (which is well measured) and the $J=1$ projection
of the $\pi\pi\to \eta\pi$ amplitude. This amplitude was evaluated
in~\cite{Descotes-Genon:2014tla} using Khuri-Treiman equations
with experimental inputs on $\eta\to3\pi$ decays.    

Let us now consider the scalar form factor  $f_0^{\pi\eta}$.
In that case, the $\ket{\piz\pip}$ contribution to 
unitarity can be neglected because it is quadratic in isospin
breaking. The important low energy scalar resonance is the
$a_0(980)$ which is known to couple strongly to the two channels
$\ket{\eta\pip}$ and $\ket{\Kzb\Kp}$.
The unitarity relation for the scalar form factor, including these two
channels, reads (see ref.~\cite{Descotes-Genon:2014tla} for details on
the derivation)
\be\lbl{unitf_0}
\ba{ll}
\im[f_0^{\eta\pip}(s)]
& =\sigma_{\eta\pip}(s)\, f_0^{\eta\pip}(s)\times
(t^{\eta\pip\to \eta\pip}_{J=0}(s))^* \\[0.3cm]
& + \sigma_{\Kzb\Kp}(s)\, \dfrac{\Delta_{\Kzb\Kp}}{\sqrt2{\Delta_{\eta\pi}}}
f_0^{\Kzb\Kp}(s)\times  (t^{\eta\pip\to\Kzb\Kp}_{J=0}(s))^* 
\ea\en
where $\sigma_{12}(s)=\theta(s-(m_1+m_2)^2)
  \sqrt{\lambda_{12}(s)}/{s} $
and a similar relation can be written for $\im[f_0^{\Kzb\Kp}]$. This
suggests a minimal representation of the scalar form factors analogous to
eq.~\rf{Kpiomnes} involving an Omn\`es matrix instead of a function,
\be\lbl{mo2x2min}
\bp
f_0^{\eta\pip}(s)\\[0.2cm]
\epsilon_{KK}f_0^{\Kzb\Kp}(s)\ep\simeq \bp
\Omega_{11}(s) &  \Omega_{12}(s) \\[0.1cm]
\Omega_{21}(s)&  \Omega_{22}(s) \ep
\bp
f_0^{\eta\pip}(0)\\[0.1cm]
\epsilon_{KK}f_0^{\Kzb\Kp}(0)\ep
\en
with\footnote{The minus sign arises because the $\Omega$ matrix is defined with
respect to isospin eigenstates and one has $\ket{\eta\pip}=-\ket{1,1}$.}
$\epsilon_{KK}=-\Delta_{\Kzb\Kp}/\sqrt2\Delta_{\eta\pi}$. The
coupled-channel Omn\`es-Muskhelishvili problem cannot be solved in
closed form in terms of phase-shifts and inelasticities as in the
one-channel case, but the equations can be solved numerically in terms
of the $T$-matrix elements~\cite{Donoghue:1990xh}. Furthermore,
no direct experimental determinations of the phase-shifts
and inelasticities has been performed for $\eta\pi$ scattering.

\section{The $\Omega$ matrix in $\gamma\gamma$ scattering and $\phi(1020)$
  radiative decay}
Cross-sections for $\gamma\gamma\to \eta\piz$ have first been measured
by the Crystal Ball collaboration at SLAC~\cite{Edwards:1981np} and at
DESY~\cite{Antreasyan:1985wx}.  Recently, high statistics measurements
have been performed by the Belle
collaboration~\cite{Uehara:2009cf}. 
The $\gamma\gamma\to \eta\piz$ process is described by helicity amplitudes
$L_{\lambda\lambda'}$ which are functions of the Mandelstam variables $s$ (the
energy squared of the $\eta\piz$ pair) and $t$, $u$.
In the energy region $\sqrt{s}\lapprox 1.4$ GeV the contributions of
the partial waves with $J >2$ to the cross-sections are negligibly small, while
in the region $\sqrt{s}\lapprox 1.1$ GeV, the $S$-wave $l_{0++}(s)$
dominates.  
Theoretical descriptions of this amplitude based on general FSI methods have
been discussed recently~\cite{Danilkin:2017lyn,Lu:2020qeo}. A closely related,
but somewhat less general approach, was used earlier in
ref.~\cite{Oller:1997yg}.  $l_{0++}(s)$ is an analytic function of $s$ except
for two cuts.

The right-hand cut
$s > (\meta+\mpi)^2$, as in the case of the form factor, is associated
with unitarity. We will consider an energy region in which two-channel
unitarity is a reasonable approximation such that one can write
\be\lbl{2gunitrels}
\ba{l}
\im[l_{0++}(s)]=\sigma_{\eta\piz}(s)\, l_{0++}(s)\,T_{11}^*(s)
              +\sigma_{K\Kbar}(s)\,k^1_{0++}(s)\,T_{12}^*(s)\\
\im[k^1_{0++}(s)]=\sigma_{\eta\piz}(s)\, l_{0++}(s)\,T_{12}^*(s)
              +\sigma_{K\Kbar}(s)\,k^1_{0++}(s)\,T_{22}^*(s)\
\ea\en
where $k^1_{0++}(s)$ is the $J=0$ $\gamma\gamma\to (K\Kbar)^{I=1}$  amplitude
and the $T$-matrix elements associated with the states $\ket{\eta\piz}$,
$\ket{K\Kbar}$ are denoted simply as $T_{ij}$.
In addition to the right-hand cut, the amplitudes $l_{0++}(s)$, $k^1_{0++}(s)$
have a left-hand cut on the negative real axis $-\infty < s \le
0$. There are no other singularities. The amplitudes must also satisfy Low's 
soft photon theorem~\cite{Low:1958sn} which states that
\be
\lim_{s\to 0} l_{0++}(s) =0,\quad
\lim_{s\to 0}\bar{k}^1_{0++}(s)=0
\en
where $\bar{k}^1_{0++}\equiv k^1_{0++}(s)-k^{1,Born}_{0++}(s)$ and
$k^{1,Born}_{0++}$ is the $I=1$ projection of the amplitude part
induced by the Kaon pole in the $t$ and $u$ channels, which coincides
with the QED Born amplitude
\be\lbl{k1Born}
k^{1,Born}_{0++}(s)=-\frac{2\sqrt2{m}^2_\Kp}{s\sigma_\Kp(s)}
\log\frac{1+\sigma_\Kp(s)}{1-\sigma_\Kp(s)}\ .
\en
and $\sigma_\Kp(s)=\sqrt{1-4\mkd/s}$. 
Knowing the location of the singularities allows one to write Cauchy
dispersive integral representations~\footnote{It is also necessary to know the
asymptotic behaviour. It can be argued~\cite{Lu:2020qeo} that $|l_{0++}(s)|$,
$|k^1_{0++}(s)|$ are bounded by $\sqrt{s}$ when $s\to\infty$.}
for $l_{0++}$ and $k^1_{0++}$. Combining
these with the unitarity relations~\rf{2gunitrels} leads to a set of
Muskhelishvili-type singular equations for the amplitudes. The
solutions, taking into account the soft photon constraints, can be expressed
in the following form in terms of the Omn\`es matrix~\cite{Lu:2020qeo}
\be\lbl{MOamplitudes}
\bp
l_{0++}(s)\\[0.15cm]
k^1_{0++}(s)\\
\ep =\bp
0\\[0.15cm]
k^{1,Born}_{0++}(s)\\
\ep  +s\bp
{\Omega_{11}(s)}&{\Omega_{12}(s)}\\[0.15cm]
{\Omega_{21}(s)}&{\Omega_{22}(s)}\\
\ep\bp
{b_l} + L_1^V(s)+ R_1^{Born}(s)\\[0.15cm]
{b_k} + L_2^V(s)+ R_2^{Born}(s)\\
\ep\ .
\en
In these equations the functions $R_i^{Born}$ are integrals involving the Born
amplitude,
\be
R_i^{Born}(s)=-\frac{s-s_A}{\pi}\int_{4\mkd}^\infty
\frac{ds'}{s'(s'-s_A)(s'-s)}\im[(\Omega^{-1})_{i2}(s')] k^{1,Born}_{0++}(s')
\en
where $s_A$ can conveniently be taken to be equal to the Adler zero of
the amplitude $l_{0++}$ ($s_A\simeq m^2_\eta$) and the functions
$L_i^V(s)$ are integrals over the left-hand cut involving
$\im[l_{0++}(s')]$, $\im[\bar{k}^1_{0++}(s')]$.  These imaginary parts
are approximated by the contributions induced by the light vector
meson poles in the $t$ (or $u$) channels, i.e.
\be
\im[l_{0++}(s')]\simeq \sum_{V=\rho,\omega,\phi} \im[l^V_{0++}(s')],\quad
\im[\bar{k}^1_{0++}(s')]\simeq \im[\bar{k}^{1,K^*}_{0++}(s')],\quad (s'\le 0)\ .
\en
As before, the contributions from the higher energy regions of the
cuts, where the approximations made are no longer accurate, are
absorbed into a set of subtraction parameters. Eq.~\rf{MOamplitudes}
involves two such parameters $b_l$, $b_k$.

The representation~\rf{MOamplitudes} for the $S$-wave $\gamma\gamma$
amplitudes, supplemented with a simple Breit-Wigner model for the $D$-waves,
was compared to the experimental data on photon-photon scattering to $\eta\pi$
cross-sections in~\cite{Danilkin:2017lyn,Lu:2020qeo}. In
ref.~\cite{Lu:2020qeo} experimental data on $\gamma\gamma\to K_SK_S, \Kp\Km$
was considered as well, which allows to constrain not only the constants
$b_l$, $b_k$ but also the $\Omega$ matrix. For that purpose, a previously
proposed six parameters model for the underlying
$T$-matrix~\cite{Albaladejo:2015aca} was used, from which $\Omega$ is computed
by solving numerically the related Muskhelishvili-Omn\`es equations.
Fig.~\fig{compexp} shows an illustrative comparison of cross sections
computed from this fitted theoretical model with the experimental ones.

Let us now consider the case where one photon has a non-vanishing virtuality
$\qd$, and label the $S$-wave amplitudes as $l_{0++}(s,\qd)$,
$k^1_{0++}(s,\qd)$. These amplitudes are analytic functions of the energy $s$
with two cuts (there are no anomalous thresholds), such that one can
write a dispersive representation in terms of the Omn\`es matrix, exactly as
before, 
\be\lbl{MOdecay}
\bp
l_{0++}(s,\qd)\\
k_{0++}(s,\qd)\ep\!=\!\bp
0\\
\tilde{k}^{1,Born}(s,\qd)\ep \!\!
+(s-\qd)
\bp\!\!
{\Omega_{11}}(s) & {\Omega_{12}}(s) \\
{\Omega_{21}}(s) & {\Omega_{22}}(s)\ep\!\!
\bp
{a_1}(\qd) + I_1^{LC}(s,\qd)+  I_1^{RC}(s,\qd)\\
{a_2}(\qd) + I_2^{LC}(s,\qd)+  I_2^{RC}(s,\qd) \ep\
\en
satisfying the soft photon theorem which now corresponds to $s=\qd$. We will
be interested in the case of timelike virtualities and $\qd > (\meta+\mpi)^2$
which corresponds experimentally to the $e^+ e^- \to \gamma^*\to
\gamma\eta\pi$ processes (and also, up to an isospin rotation, to the vector
current part of the radiative $\tau$ decays $\tau^+ \to \gamma\eta\pi^+\nu,
\gamma\bar{K}^0K^+\nu$). Some care must be taken in order to properly utilize
eq.~\rf{MOdecay} because the pole singularity of the Born amplitude at $s=\qd$
occurs within the range of integration in the integrals $I^{RC}_i$. It is
convenient to separate this singular part (labelled
$\tilde{k}^{1,Born}$ in~\rf{MOdecay} )and
perform the related integrations analytically. Furthermore, the left-hand cut
now has complex components as well as a real component which partly overlaps
with the unitarity cut (which leads to a violation of the Fermi-Watson phase
theorem).  Details on how to properly compute the left-cut integrals
$I_i^{LC}$ can be found in~\cite{Moussallam:2021dpk}). In practice,
eqs.~\rf{MOdecay} where implemented with $\qd\simeq m^2_\phi$ enabling to
determine the $\phi(1020)\to \gamma\eta\pi$ decay amplitude. As in the $\qd=0$
case, the discontinuities across the left-hand cut components are modelled
based on the contributions from the light vector mesons $\rho$, $\omega$,
$\phi$ and $K^*$. Fig.~\fig{compexp} shows that a rather good description of
the accurate experimental data~\cite{Aloisio:2002bsa,Ambrosino:2009py} can be
obtained with a two parameter fit, using the $\Omega$ matrix as determined
previously from the $\gamma\gamma$ scattering fit.

\begin{figure}[h]
\hspace*{-0.8cm}
\includegraphics[width=0.36\textwidth]{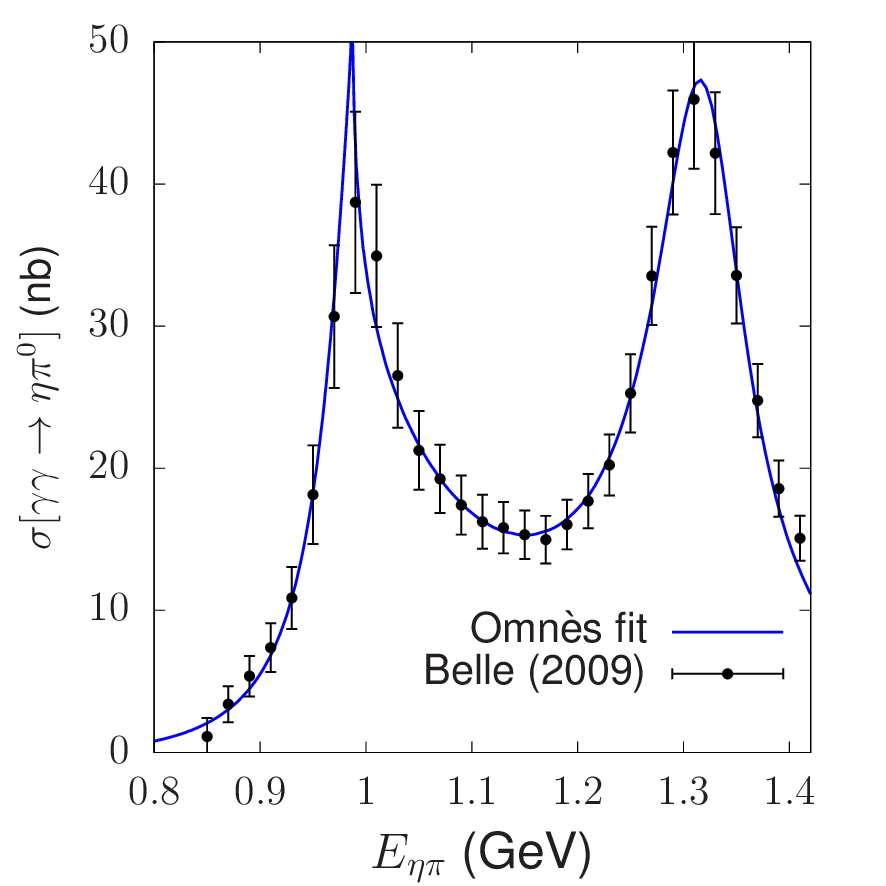}%
\includegraphics[width=0.36\textwidth]{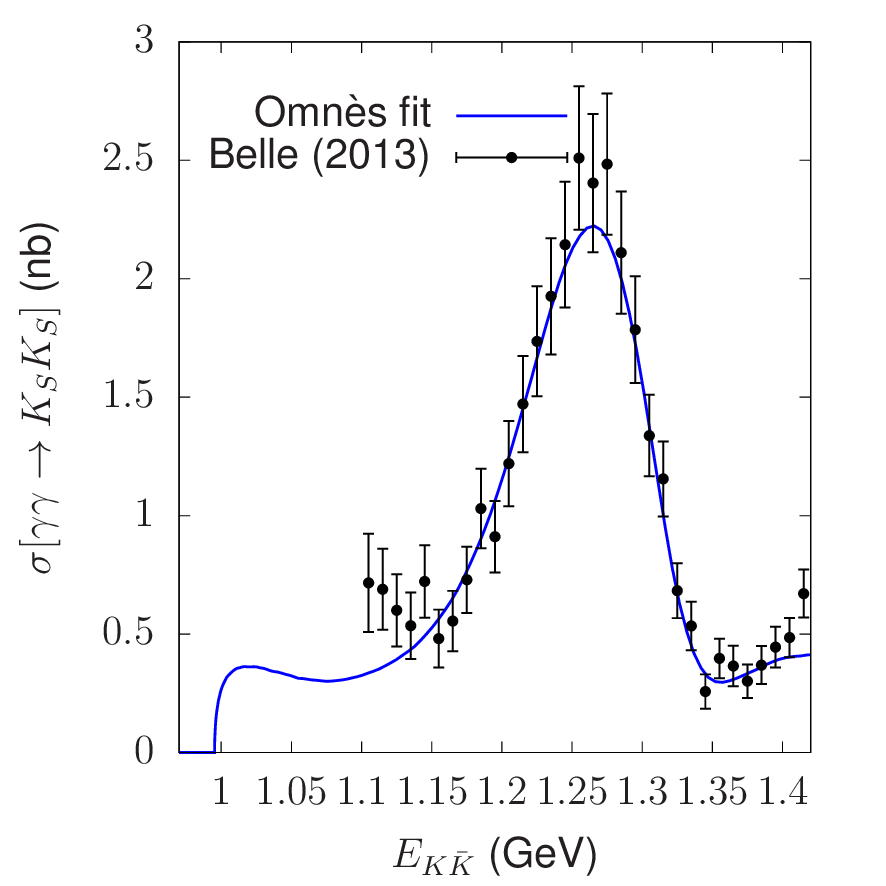}%
\includegraphics[width=0.36\textwidth]{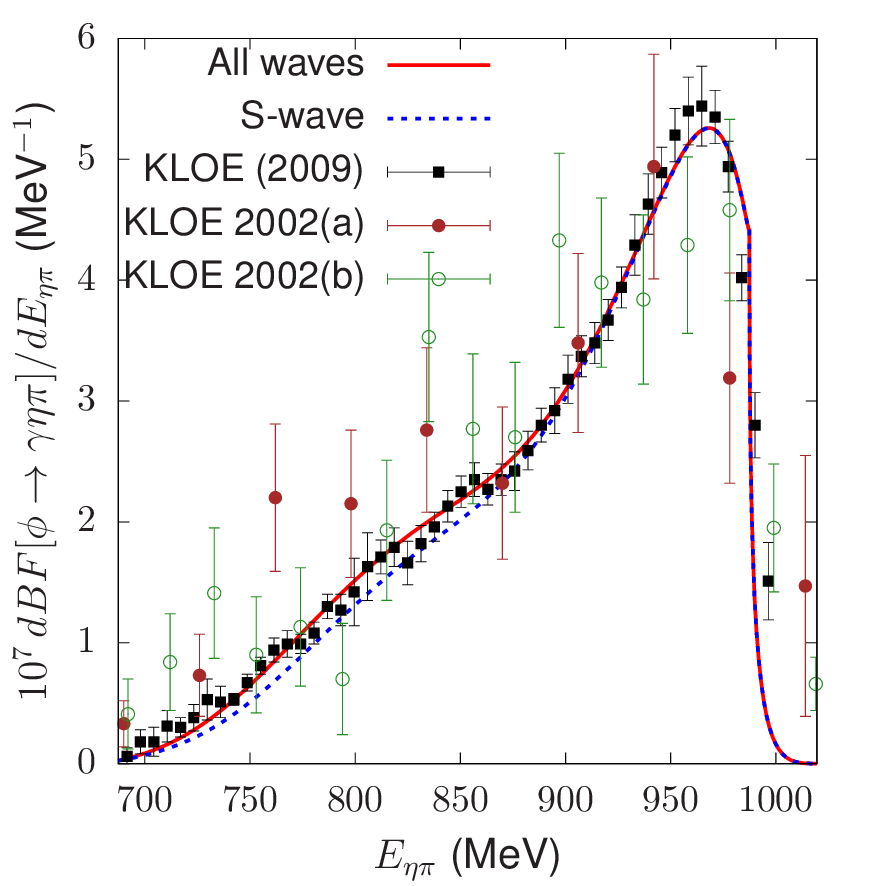}
\caption{\small
Comparison of experimental data with Omn\`es-based $S$-wave amplitude
models, left: $\gamma\gamma\to\pi\eta$ cross-sections, middle:
$\gamma\gamma\to K_S K_S$ cross-sections, right: $\phi\to \gamma\eta\pi$
energy distribution.} 
\lblfig{compexp}
\end{figure}
\section{Chiral inputs and results}
In the minimal modelling, once the $\Omega$ matrix is known the scalar
form factors are given in terms of their values at $s=0$. Since
$f_0^{\Kzb\Kp}(0)$ appears multiplied by an isospin
breaking factor, we can simply set $f_0^{\Kzb\Kp}(0)=1$, because the chiral
corrections are $O(m_d-m_u,e^2)$. The chiral expansion of
$f_0^{\eta\pip}(0)$ takes the following form~\cite{Neufeld:1994eg}
\be\lbl{fetapi0chir}
f_0^{\eta\pip}(0)=\epsilon\times(1+O(p^2)) +\frac{O(e^2p^2)}{\metad-\mpid},\quad
\epsilon=\frac{\sqrt3(m_d-m_u)}{4(m_s-\hat{m})}
\en
with $\hat{m}=(m_d+m_u)/2$. The value of $\epsilon$
from the FLAG revue~\cite{Aoki:2021kgd} is
$\epsilon=1.137(45)\cdot10^{-2}$. The $O(p^2)$ correction involves the
chiral couplings $L_7$, $L_8$ which can be computed using the chiral
expansions of the pseudo-scalar meson masses at NLO and the value the quark
mass ratio $m_s/\hat{m}$~\cite{Gasser:1984gg}, which is known from lattice
QCD simulations: $m_s/\hat{m}=27.42(12)$ (see~\cite{Aoki:2021kgd}). The
couplings which appear in the $O(e^2p^2)$ part in~\rf{fetapi0chir} are not
precisely known, they can be estimated approximately using resonance
saturation modelling~\cite{Baur:1996ya,Ananthanarayan:2004qk}. Finally, from
this NLO chiral expansion, one obtains
\be\lbl{fetapi0numer}
\left.f_0^{\eta\pip}(0)\right\vert_{\chi{PT}}=(1.40\pm0.14)\cdot10^{-2}\ ,
\en
assuming a 50\% error on the chiral corrections. An alternative
estimate can be performed based on a relation, which is exact at NLO,
with the $K^+_{l3}$ and $K^0_{l3}$ form factors ratio
$\Delta_{SU(2)}$ $\equiv{f_0^{\Kp\piz}(0)}/{f_0^{\Kz\pim}(0)}-1$~\cite{Neufeld:1994eg}.
The experimental value $\Delta_{SU(2)}$
$=(2.73\pm0.41)\cdot10^{-2}$~\cite{Antonelli:2010yf} gives
$f_0^{\eta\pip}(0)\vert_{\chi{PT}+K_{l3}}$
$=(1.51\pm0.24)\cdot10^{-2}$ which is compatible with
eq.~\rf{fetapi0numer}. Using the values at $s=0$ from $\chi$pt the
result for $f_0^{\eta\pi}(s)$ from the minimal Omn\`es representation
is illustrated in Fig~\fig{tauresults} (solid black line on the left
plot). Obviously, this minimal model cannot be very precise. In fact,
computing the derivatives of the form factors at $s=0$ one finds
substantial deviations from the $O(p^4)$ $\chi$pt values:
$\dot{f}_0^{\eta\pi}(0)\vert_{min}/
\dot{f}_0^{\eta\pi}(0)\vert_{\chi{pt}}\sim0.5$,   
$\dot{f}_0^{K\Kbar}(0)\vert_{min}/
\dot{f}_0^{K\Kbar}(0)\vert_{\chi{pt}}\sim1.6$.  
The values of the derivatives can be corrected by replacing
$f_0^{12}(0)$ in eq.~\rf{mo2x2min} by a linear polynomial
$f_0^{12}(0)(1+\lambda_{12} s)$. Physically, it is plausible that
two-channel unitarity is insufficient in the region of the $a_0(1450)$
resonance. Assuming an effective third channel, this suggests the
following simple improvement over the minimal model
\be
f_0^{\eta\pi}(s)= \left.f_0^{\eta\pi}(s)\right\vert_{min}
+\Omega_{13}(s)\lambda_{13},\quad
\Omega_{13}(s)\simeq\frac{s }{m^2_{a_0'}-s-i m_{a_0'} \Gamma_{a_0'} }
\en
in which the parameter $\lambda_{13}$ is adjusted such that the $\chi$pt
derivative at $s=0$ is reproduced. The effects of these various ways to modify
the derivatives at $s=0$ are illustrated in Fig.~\fig{tauresults}.

\begin{figure}[h]
\hfill\includegraphics[width=0.40\linewidth]{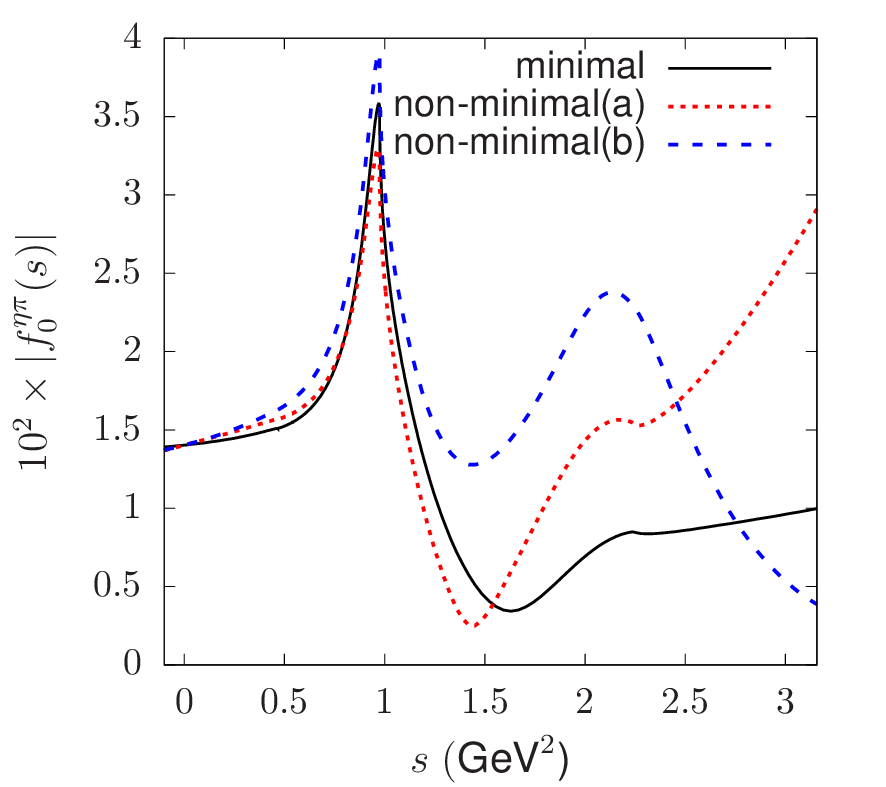}%
\hfill\includegraphics[width=0.40\linewidth]{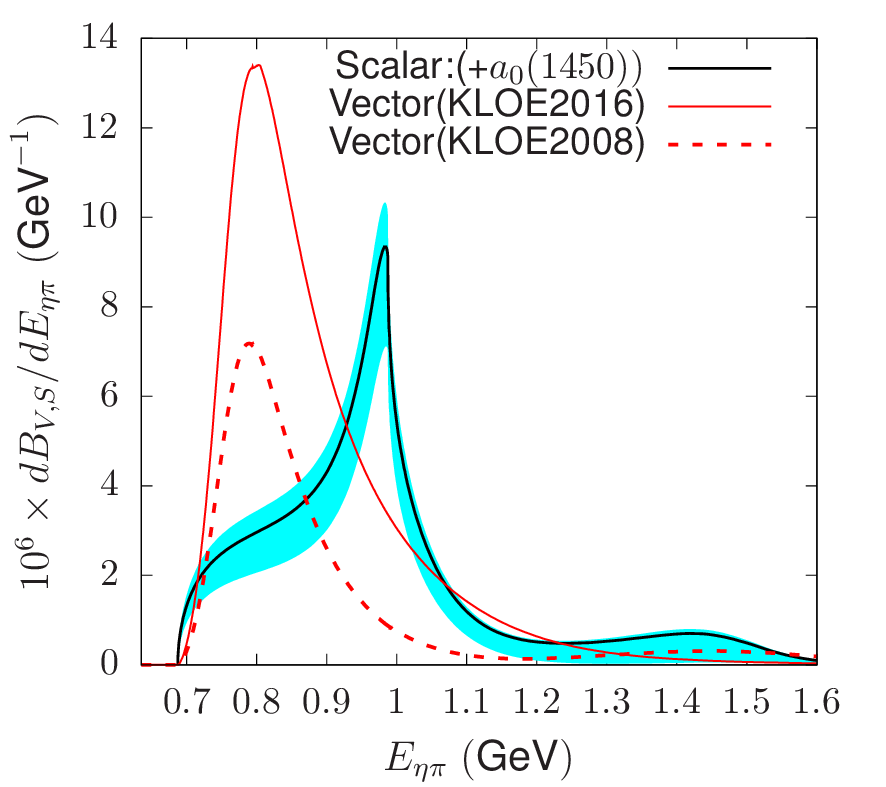}\hfill
\caption{\small Left plot: absolute value of the scalar form factor
  $f_0^{\eta\pi}$ from the minimal two-channel Omn\`es model (black solid line)
  and from non-minimal models a) imposing the value of the derivative
  at $s=0$ via a linear polynomial, b) imposing this value via a third channel. 
The
  right plot shows the energy distribution of the $\tau\to \eta\pi\nu$ branching
  fractions. The black solid line corresponds to the scalar form factor and
  blue error band is generated by varying both the input values at $s=0$ and
  those of the derivatives. The red lines correspond to the vector form
  factor (see text).}
\lblfig{tauresults}
\end{figure}

The energy dependence of the $\tau\to\eta\pi\nu$ branching fraction
associated to the scalar form factor is shown in Fig.~\fig{tauresults}
(right). Also shown for comparison is the branching fraction
associated with the vector form factor. The dotted red line is the
result from ref.~\cite{Descotes-Genon:2014tla} in which the amplitude
$t_{J=1}^{\piz\pip\to\eta\pip}$ is determined using $\eta\to3\pi$
inputs from ref.~\cite{Ambrosino:2008ht}. The solid red line is an
update which uses the more recent results
from~\cite{Anastasi:2016qvh}. One sees a substantial difference
between these two evaluations which gives a measure of the error.
Finally, integrating over the $\pi\eta$ energy we find the following
results for the branching fractions
\be\lbl{BFresults}
BF_S= (1.8\pm 0.7)\cdot10^{-6}, \quad BF_V= (2.6\pm 1.30)\cdot10^{-6}\ .
\en

\section{Conclusions}
We have proposed a determination of the scalar form factor component
of the $\tau\to\eta\pi\nu$ amplitude based on a two-channel Omn\`es
matrix constrained by photon-photon scattering data and shown to be
compatible with $\phi$ radiative decay data. The result for the
branching fraction $BF_S$ (eq.~\rf{BFresults}) should be more precise
than the one given previously
in~\cite{Descotes-Genon:2014tla}~\footnote{In that reference and idea
  proposed in ref.~\cite{Yndurain:2003vk} was followed, which consists in
  making estimates of the behavior of the phase of the form factor in
  the inelastic region using the QCD asymptotic constraint.}. 
We have also presented an update on the branching fraction generated
by the vector form factor.  If our results are correct, then the
forthcoming search for this rare $\tau$ decay mode at Belle II might
not be able to clearly see it. However, these new measurements should
be able to rule out some of the theoretical models. 

\section*{Acknowledgements}
This work is supported  by the European Union's Horizon2020 research
and innovation programme (HADRON-2020) under the Grant Agreement n$^\circ$
824093.

\nolinenumbers


\begin{thebibliography}{10}
\providecommand{\url}[1]{\texttt{#1}}
\providecommand{\urlprefix}{URL }
\expandafter\ifx\csname urlstyle\endcsname\relax
  \providecommand{\doi}[1]{doi:\discretionary{}{}{}#1}\else
  \providecommand{\doi}{doi:\discretionary{}{}{}\begingroup
  \urlstyle{rm}\Url}\fi
\providecommand{\eprint}[2][]{\url{#2}}

\bibitem{Weinberg:1958ut}
S.~Weinberg,
\newblock \emph{{Charge symmetry of weak interactions}},
\newblock Phys.Rev. \textbf{112}, 1375 (1958),
\newblock \doi{10.1103/PhysRev.112.1375}.

\bibitem{Bramon:1987zb}
A.~Bramon, S.~Narison and A.~Pich,
\newblock \emph{{The $\tau\to \nu_\tau\eta\pi $ Process in and beyond QCD}},
\newblock Phys.Lett. \textbf{B196}, 543 (1987),
\newblock \doi{10.1016/0370-2693(87)90817-3}.

\bibitem{Nussinov:2008gx}
S.~Nussinov and A.~Soffer,
\newblock \emph{{Estimate of the branching fraction $\tau\to\eta\pi^-
  \nu_\tau$, the $a_0(980)$, and non-standard weak interactions}},
\newblock Phys.Rev. \textbf{D78}, 033006 (2008),
\newblock \doi{10.1103/PhysRevD.78.033006},
\newblock \eprint{0806.3922}.

\bibitem{Garces:2017jpz}
E.~A. Garc\'es, M.~Hern\'andez~Villanueva, G.~L\'opez~Castro and P.~Roig,
\newblock \emph{{Effective-field theory analysis of the $\tau^- \to
  \eta^{(\prime)} \pi^- \nu_\tau$ decays}},
\newblock JHEP \textbf{12}, 027 (2017),
\newblock \doi{10.1007/JHEP12(2017)027},
\newblock \eprint{1708.07802}.

\bibitem{Derrick:1987sp}
M.~Derrick \emph{et~al.},
\newblock \emph{{Evidence for the Decay $\tau^+ \to \pi^+ \eta
  \bar{\tau}$-neutrino}},
\newblock Phys. Lett. B \textbf{189}, 260 (1987),
\newblock \doi{10.1016/0370-2693(87)91308-6}.

\bibitem{Hayasaka:2009zz}
K.~Hayasaka,
\newblock \emph{{Second class current in $\tau\to \pi \eta \nu$ analysis and
  measurement of $\tau \to h h' h'' \nu$ from Belle: electroweak physics from
  Belle}},
\newblock PoS \textbf{EPS-HEP2009}, 374 (2009),
\newblock \doi{10.22323/1.084.0374}.

\bibitem{delAmoSanchez:2010pc}
P.~del Amo~Sanchez \emph{et~al.},
\newblock \emph{{Studies of $\tau^- \to \eta K^-\nu_\tau$ and $\tau^- \to \eta
  \pi^- \nu_\tau$ at BaBar and a search for a second-class current}},
\newblock Phys.Rev. \textbf{D83}, 032002 (2011),
\newblock \doi{10.1103/PhysRevD.83.032002},
\newblock \eprint{1011.3917}.

\bibitem{Ogawa:2020iwi}
K.~Ogawa, M.~H. Villanueva and K.~Hayasaka,
\newblock \emph{{Search for second-class currents with the $\tau$ decay into
  $\pi\eta\nu$}},
\newblock PoS \textbf{Beauty2019}, 061 (2020),
\newblock \doi{10.22323/1.377.0061}.

\bibitem{Bhattacharya:2011qm}
T.~Bhattacharya, V.~Cirigliano, S.~D. Cohen, A.~Filipuzzi, M.~Gonzalez-Alonso,
  M.~L. Graesser, R.~Gupta and H.-W. Lin,
\newblock \emph{{Probing Novel Scalar and Tensor Interactions from (Ultra)Cold
  Neutrons to the LHC}},
\newblock Phys. Rev. D \textbf{85}, 054512 (2012),
\newblock \doi{10.1103/PhysRevD.85.054512},
\newblock \eprint{1110.6448}.

\bibitem{Tisserant:1982fc}
S.~Tisserant and T.~Truong,
\newblock \emph{{$\tau\to \delta\nu$ Decay induced by Light Quark Mass
  Difference}},
\newblock Phys.Lett. \textbf{B115}, 264 (1982),
\newblock \doi{10.1016/0370-2693(82)90659-1}.

\bibitem{Pich:1987qq}
A.~Pich,
\newblock \emph{{'Anomalous' $\eta$ Production in Tau Decay}},
\newblock Phys.Lett. \textbf{B196}, 561 (1987),
\newblock \doi{10.1016/0370-2693(87)90821-5}.

\bibitem{Neufeld:1994eg}
H.~Neufeld and H.~Rupertsberger,
\newblock \emph{{Isospin breaking in chiral perturbation theory and the decays
  $\eta \to \pi l \nu$ and $\tau \to \eta \pi \nu$}},
\newblock Z.Phys. \textbf{C68}, 91 (1995),
\newblock \doi{10.1007/BF01579808}.

\bibitem{Paver:2010mz}
N.~Paver and Riazuddin,
\newblock \emph{{On meson dominance in the `second class' $\tau \to \eta \pi
  \nu_\tau$ decay}},
\newblock Phys.Rev. \textbf{D82}, 057301 (2010),
\newblock \doi{10.1103/PhysRevD.82.057301},
\newblock \eprint{1005.4001}.

\bibitem{Volkov:2012be}
M.~Volkov and D.~Kostunin,
\newblock \emph{{The decays $\rho^{-}\to\eta\pi^{-}$ and
  $\tau^{-}\to\eta(\eta')\pi^{-}\nu$ in the NJL model}},
\newblock Phys.Rev. \textbf{D86}, 013005 (2012),
\newblock \doi{10.1103/PhysRevD.86.013005},
\newblock \eprint{1205.3329}.

\bibitem{Descotes-Genon:2014tla}
S.~Descotes-Genon and B.~Moussallam,
\newblock \emph{{Analyticity of $\eta \pi $ isospin-violating form factors and
  the $\tau \rightarrow \eta \pi \nu $ second-class decay}},
\newblock Eur. Phys. J. \textbf{C74}, 2946 (2014),
\newblock \doi{10.1140/epjc/s10052-014-2946-8},
\newblock \eprint{1404.0251}.

\bibitem{Escribano:2016ntp}
R.~Escribano, S.~Gonzalez-Solis and P.~Roig,
\newblock \emph{{Predictions on the second-class current decays
  $\tau^{-}\to\pi^{-}\eta^{(\prime)}\nu_{\tau}$}},
\newblock Phys. Rev. D \textbf{94}(3), 034008 (2016),
\newblock \doi{10.1103/PhysRevD.94.034008},
\newblock \eprint{1601.03989}.

\bibitem{Barton:1965}
G.~Barton,
\newblock \emph{Introduction to dispersion techniques in field theory},
\newblock Lecture notes and supplements in physics. W.A. Benjamin, New York
  (1965).

\bibitem{Omnes:1958hv}
R.~Omn\`es,
\newblock \emph{{On the Solution of certain singular integral equations of
  quantum field theory}},
\newblock Nuovo Cim. \textbf{8}, 316 (1958),
\newblock \doi{10.1007/BF02747746}.

\bibitem{Donoghue:1990xh}
J.~F. Donoghue, J.~Gasser and H.~Leutwyler,
\newblock \emph{{The Decay of a Light Higgs Boson}},
\newblock Nucl.Phys. \textbf{B343}, 341 (1990),
\newblock \doi{10.1016/0550-3213(90)90474-R}.

\bibitem{Edwards:1981np}
C.~Edwards \emph{et~al.},
\newblock \emph{{Production of $\pi^0 \pi^0$ and $\pi^0 \eta$ in Photon -
  Photon Collisions}},
\newblock Phys. Lett. B \textbf{110}, 82 (1982),
\newblock \doi{10.1016/0370-2693(82)90957-1}.

\bibitem{Antreasyan:1985wx}
D.~Antreasyan \emph{et~al.},
\newblock \emph{{Formation of $\delta(980)$ and $A_2(1320)$ in Photon-photon
  Collisions}},
\newblock Phys. Rev. \textbf{D33}, 1847 (1986),
\newblock \doi{10.1103/PhysRevD.33.1847}.

\bibitem{Uehara:2009cf}
S.~Uehara \emph{et~al.},
\newblock \emph{{High-statistics study of eta pi0 production in two-photon
  collisions}},
\newblock Phys.Rev. \textbf{D80}, 032001 (2009),
\newblock \doi{10.1103/PhysRevD.80.032001},
\newblock \eprint{0906.1464}.

\bibitem{Danilkin:2017lyn}
I.~Danilkin, O.~Deineka and M.~Vanderhaeghen,
\newblock \emph{{Theoretical analysis of the $\gamma\gamma \to \pi^0 \eta$
  process}},
\newblock Phys. Rev. \textbf{D96}(11), 114018 (2017),
\newblock \doi{10.1103/PhysRevD.96.114018},
\newblock \eprint{1709.08595}.

\bibitem{Lu:2020qeo}
J.~Lu and B.~Moussallam,
\newblock \emph{{The $\pi \eta $ interaction and $a_0$ resonances in
  photon\textendash{}photon scattering}},
\newblock Eur. Phys. J. C \textbf{80}(5), 436 (2020),
\newblock \doi{10.1140/epjc/s10052-020-7969-8},
\newblock \eprint{2002.04441}.

\bibitem{Oller:1997yg}
J.~A. Oller and E.~Oset,
\newblock \emph{{Theoretical study of the gamma gamma $\to$ meson - meson
  reaction}},
\newblock Nucl. Phys. \textbf{A629}, 739 (1998),
\newblock \doi{10.1016/S0375-9474(97)00649-0},
\newblock \eprint{hep-ph/9706487}.

\bibitem{Low:1958sn}
F.~Low,
\newblock \emph{{Bremsstrahlung of very low-energy quanta in elementary
  particle collisions}},
\newblock Phys.Rev. \textbf{110}, 974 (1958),
\newblock \doi{10.1103/PhysRev.110.974}.

\bibitem{Albaladejo:2015aca}
M.~Albaladejo and B.~Moussallam,
\newblock \emph{{Form factors of the isovector scalar current and the $\eta
  \pi$ scattering phase shifts}},
\newblock Eur. Phys. J. C \textbf{75}(10), 488 (2015),
\newblock \doi{10.1140/epjc/s10052-015-3715-z},
\newblock \eprint{1507.04526}.

\bibitem{Moussallam:2021dpk}
B.~Moussallam,
\newblock \emph{{Revisiting $\gamma ^*\rightarrow \gamma \pi ^0\eta $ near the
  $\phi (1020)$ using analyticity and the left-cut structure}},
\newblock Eur. Phys. J. C \textbf{81}(11), 993 (2021),
\newblock \doi{10.1140/epjc/s10052-021-09772-8},
\newblock \eprint{2107.14147}.

\bibitem{Aloisio:2002bsa}
A.~Aloisio \emph{et~al.},
\newblock \emph{{Study of the decay $\phi\to \eta \pi^0 \gamma$ with the KLOE
  detector}},
\newblock Phys.Lett. \textbf{B536}, 209 (2002),
\newblock \doi{10.1016/S0370-2693(02)01821-X},
\newblock \eprint{hep-ex/0204012}.

\bibitem{Ambrosino:2009py}
F.~Ambrosino \emph{et~al.},
\newblock \emph{{Study of the $a_0(980)$ meson via the radiative decay $\phi\to
  \eta \pi^0 \gamma$ with the KLOE detector}},
\newblock Phys.Lett. \textbf{B681}, 5 (2009),
\newblock \doi{10.1016/j.physletb.2009.09.022},
\newblock \eprint{0904.2539}.

\bibitem{Aoki:2021kgd}
Y.~Aoki \emph{et~al.},
\newblock \emph{{FLAG Review 2021}}  (2021),
\newblock \eprint{2111.09849}.

\bibitem{Gasser:1984gg}
J.~Gasser and H.~Leutwyler,
\newblock \emph{{Chiral Perturbation Theory: Expansions in the Mass of the
  Strange Quark}},
\newblock Nucl.Phys. \textbf{B250}, 465 (1985),
\newblock \doi{10.1016/0550-3213(85)90492-4}.

\bibitem{Baur:1996ya}
R.~Baur and R.~Urech,
\newblock \emph{{Resonance contributions to the electromagnetic low-energy
  constants of chiral perturbation theory}},
\newblock Nucl. Phys. B \textbf{499}, 319 (1997),
\newblock \doi{10.1016/S0550-3213(97)00348-9},
\newblock \eprint{hep-ph/9612328}.

\bibitem{Ananthanarayan:2004qk}
B.~Ananthanarayan and B.~Moussallam,
\newblock \emph{{Four-point correlator constraints on electromagnetic chiral
  parameters and resonance effective Lagrangians}},
\newblock JHEP \textbf{06}, 047 (2004),
\newblock \doi{10.1088/1126-6708/2004/06/047},
\newblock \eprint{hep-ph/0405206}.

\bibitem{Antonelli:2010yf}
M.~Antonelli, V.~Cirigliano, G.~Isidori, F.~Mescia, M.~Moulson \emph{et~al.},
\newblock \emph{{An Evaluation of $|V_us|$ and precise tests of the Standard
  Model from world data on leptonic and semileptonic kaon decays}},
\newblock Eur.Phys.J. \textbf{C69}, 399 (2010),
\newblock \doi{10.1140/epjc/s10052-010-1406-3},
\newblock \eprint{1005.2323}.

\bibitem{Ambrosino:2008ht}
F.~Ambrosino \emph{et~al.},
\newblock \emph{{Determination of $\eta \to \pi^+ \pi^- \pi^0$ Dalitz plot
  slopes and asymmetries with the KLOE detector}},
\newblock JHEP \textbf{0805}, 006 (2008),
\newblock \doi{10.1088/1126-6708/2008/05/006},
\newblock \eprint{0801.2642}.

\bibitem{Anastasi:2016qvh}
A.~Anastasi \emph{et~al.},
\newblock \emph{{Precision measurement of the $\eta\to\pi^+\pi^-\pi^0$ Dalitz
  plot distribution with the KLOE detector}},
\newblock JHEP \textbf{05}, 019 (2016),
\newblock \doi{10.1007/JHEP05(2016)019},
\newblock \eprint{1601.06985}.

\bibitem{Yndurain:2003vk}
F.~Yndur{\'a}in,
\newblock \emph{{The Quadratic scalar radius of the pion and the mixed pi-k
  radius}},
\newblock Phys.Lett. \textbf{B578}, 99 (2004),
\newblock \doi{10.1016/j.physletb.2003.10.037, 10.1016/j.physletb.2004.03.001},
\newblock \eprint{hep-ph/0309039}.

\end{thebibliography}

\end{document}